\documentclass[article]{revtex4}

\usepackage{amsmath,amssymb,amsthm,amsfonts,mathrsfs,bm,verbatim}
\usepackage{graphicx,subfigure}

\usepackage[justification=centering]{caption}
\usepackage[colorlinks=true, allcolors=blue]{hyperref}

\def\be{\begin{equation}}
\def\ee{\end{equation}}
\def\ba{\begin{eqnarray}}\def\bea{\begin{eqnarray}}
\def\ea{\end{eqnarray}}  \def\eea{\end{eqnarray}}

\def\l{\lambda}\def\o{\omega}\def\D{\Delta}
\def\rp{r_+}
\def\rm{r_-}

\begin{document}
\title{Superradiant stability of the Kerr black holes}
\author{Jia-Hui Huang}
\author{Wen-Xiang Chen}
\author{Zi-Yang Huang}
\affiliation{Institute of quantum matter,\\ School of Physics and Telecommunication Engineering,\\
South China Normal University,Guangzhou 510006,China}
\author{Zhan-Feng Mai}
\email{zhanfeng.mai@gmail.com}
\affiliation{Center for Joint Quantum Studies, School of Science, Tianjin University, Tianjin 300350, China}

\begin{abstract}
We study the superradiant stability of the system of a Kerr black hole and a massive scalar perturbation.
It was proved previously that this system is superradiantly stable when $\mu\geq \sqrt{2}m\Omega_H$, where $\mu$ is the
proper mass of the scalar, $m$ is the azimuthal number of the scalar mode, and $\Omega_H$ is the angular
velocity of the Kerr black hole horizon. Our study is a complementary work of this result. We analytically prove that
in the complementary parameter region $\mu<\sqrt{2}m\Omega_H$, when the parameters of scalar perturbation and
Kerr black hole satisfy two simple inequalities, $\omega<\frac{\mu}{\sqrt{2}}$,~ $\frac{r_-}{r_+}<0.802$, the system is also superradiantly stable.
\end{abstract}
\maketitle

\section{{Introduction}}
Black holes are important and peculiar objects predicted by general relativity. Aspects of black hole physics have been studied extensively. One interesting
phenomenon is the superradiant scattering of black holes\cite{Manogue1988,Greiner1985,Cardoso2004,Brito:2015oca,Brito:2014wla}, e.g., when a charged bosonic wave is impinging upon a charged rotating black hole, the wave is amplified by the black hole if the wave frequency $\omega$ obeys
  \begin{equation}\label{superRe}
   \omega < m\Omega_H  + e\Phi,
  \end{equation}
where $e$ and $m$ are the charge and azimuthal number of the bosonic wave mode, $\Omega_H$ is the angular velocity of black hole horizon and $\Phi$ is the electromagnetic potential of the black hole horizon. This amplification is the superradiant scattering, which was studied long time ago \cite{P1969,Ch1970,M1972,Ya1971,Bardeen1972,Bekenstein1973,Damour:1976kh}, and has broad applications in various areas of physics(for a recent review, see\cite{Brito:2015oca}). Through the superradiant process, the rotational energy or electromagnetic energy of a black hole can be extracted. Due to the existence of superradiant modes, a black hole bomb mechanism was proposed by Press and Teukolsky\cite{PTbomb}. If there is a mirror between the black hole horizon and space infinity, the amplified wave can be scattered back and forth and grows exponentially, which leads to the superradiant instability of the background black hole geometry \cite{Cardoso:2004nk,Herdeiro:2013pia,Degollado:2013bha}.  Superradiant (in)stability of various kinds of black holes have been studied extensively in the literature.

For charged Reissner-Nordstrom(RN) black holes, it has been proved that they are superradiantly stable against charged massive scalar perturbation\cite{Hod:2013eea,Huang:2015jza,Hod:2015hza,DiMenza:2014vpa}. The reason is that when the superradiant modes exist in such a system of a RN black hole with a charged massive scalar wave, there is no effective trapping potential/mirror outside the black hole horizon, which reflects the
superradiant modes back and forth \cite{Hod:2015hza,Huang:2015jza}. However,
when  a mirror or a cavity is imposed outside a charged RN black hole horizon, this black hole is superradiantly unstable in certain parameter spaces \cite{Herdeiro:2013pia,Li:2014gfg,Degollado:2013bha,Sanchis-Gual:2015lje,Fierro:2017fky}. Charged black holes in curved backgrounds, such as anti-de Sitter/de Sitter(AdS/dS) space, are proved to be superradiantly unstable because these backgrounds provide  natural mirror-like boundary conditions\cite{Wang:2014eha,Bosch:2016vcp,Huang:2016zoz,Gonzalez:2017shu,Zhu:2014sya}. There is a similar case for stringy RN black holes.
 The stringy RN black hole is shown to be  superradiantly stable against charged massive scalar perturbation\cite{Li:2013jna}. But if a
mirror is introduced, superradiant modes are supported and the stringy RN black hole becomes superradiantly unstable \cite{Li:2014xxa,Li:2014fna,Li:2015mqa}.
It is also found that extra coupling between the scalar field and the gravity can result in superradiant instability of RN/RN-AdS black holes \cite{Kolyvaris:2018zxl,Abdalla:2019irr}.

For rotating Kerr black holes,  if the incoming scalar perturbation has a nonzero mass, this mass term will act as a natural mirror and lead to superradiant instability of Kerr black holes  when the parameters of the Kerr black holes and the scalar fields are in certain  parameter spaces\cite{Strafuss:2004qc,Konoplya:2006br,Cardoso:2011xi,Dolan:2012yt,Hod:2012zza,Hod:2014pza,Aliev:2014aba,Hod:2016iri,Degollado:2018ypf}. Beyond the massive scalar perturbations, superradiant instability of Kerr black holes that are impinged upon by a massive vector field is also discussed \cite{East:2017ovw,East:2017mrj}.
The superradiant instability of rotating black holes in curved space, such as Kerr-AdS black holes, has also been reported
\cite{Cardoso:2004hs,Cardoso:2013pza,Zhang:2014kna,Delice:2015zga,Aliev:2015wla,Wang:2015fgp,Ferreira:2017tnc}.

Although there has been so much study on superradiance of rotating black holes, even Kerr black hole is not investigated thoroughly. In \cite{Hod:2012zza}, the author proved that a Kerr black hole is superradiantly stable under massive scalar perturbation when
\bea\label{fanwei1}
 \mu \geq \sqrt{2}m\Omega_H,
\eea
where $\mu$ and $m$ are the mass and azimuthal number of the incident scalar wave and $\Omega_H$ is the angular velocity of the Kerr black hole horizon. In a following work \cite{Hod:2016iri}, a stronger bound on the stability regime of the  Kerr-black-hole-massive-scalar system was derived.

In order for a complete discussion of the superradiant stability of Kerr-black-hole-massive-scalar system, it is interesting to investigate the superradiance property of the system in the complementary parameter spaces  of the previous result in \cite{Hod:2012zza}. Explicitly, in the present paper, we will find the parameter regions of a Kerr black hole and a massive scalar perturbation system with $\mu<\sqrt{2}m\Omega_H$, where the system is superradiantly stable.
The scattering of massive scalar field in a Kerr black hole background can be described by a Schrodinger-like radial equation with an effective potential. In order for the instability of the system, there are two key conditions should be satisfied. One is that the parameters of the black hole and the scalar field should satisfy superradiance condition, i.e., there exist superradiance modes. The second one is that there is mirror-like condition outside the horizon of the black hole. In our case, if there is a potential well outside the black hole horizon, it will act as a reflecting mirror and may lead to instability of the system. So when there is no potential well outside the black hole horizon for the superradiance modes, the Kerr-black-hole-massive-scalar system will be superradiantly stable.

 The paper is organized as follows. In Section 2, we provide a simple introduction of the Kerr-black-hole-massive-scalar system and the angular part of the equation of motion. In Section 3, we derive the Schrodinger-like radial equation and the effective potential for the scalar perturbation in Kerr background. This effective potential is the main object as mentioned above. In Section 4, we derive the superradiantly stable parameter regions for the system based on the effective potential.
 Finally, we give a summary and discussion in Section 5.

\section{Kerr black hole and a massive scalar perturbation}
Kerr black hole describes a stationary and axially symmetric spacetime geometry. The metric of the 4-dimensional  Kerr black hole in Boyer-Lindquist coordinates is \cite{Chand1983,kerr1963,Bredberg:2009pv}(we take $G=\hbar=c=1$)
\begin{equation}
d{{s}^{2}}=\frac{\Delta }{{{\rho }^{2}}}{{(dt-a{{\sin }^{2}}\theta d\phi )}^{2}}+\frac{{{\rho }^{2}}}{\Delta }d{{r}^{2}}+{{\rho }^{2}}d{{\theta }^{2}}+\frac{{{\sin }^{2}}\theta }{{{\rho }^{2}}}{{[adt-({{r}^{2}}+{{a}^{2}})d\phi ]}^{2}},
\end{equation}
where
\begin{equation}
\Delta ={{r}^{2}}-2Mr+{{a}^{2}},~~{{\rho }^{2}}={{r}^{2}}+{{a}^{2}}{{\cos }^{2}}\theta .
\end{equation}
$M$ is the mass of the black hole and $a$ is the angular momentum per unit mass of the black hole. The inner and outer horizons of the Kerr black hole are
\begin{equation}\label{rprm-a}
{{r}_{\pm }}=M\pm \sqrt{{{M}^{2}}-{{a}^{2}}},
\end{equation}and it is obvious that
\begin{equation}
{{r}_{+}}+{{r}_{-}}=2M,~~{{r}_{+}}{{r}_{-}}={{a}^{2}}.
\end{equation}
The angular velocity of the Kerr black hole horizon is
\bea\label{ome-a}
\Omega_H=\frac{a}{r_+^2+a^2}.
\eea

The dynamics of a massive scalar field perturbation $\Psi$ with proper mass $\mu$ in the Kerr black hole background is described by the covariant Klein-Gordon equation
\begin{equation}\label{KG}
({{\nabla }^{\nu }}{{\nabla }_{\nu }}-{{\mu }^{2}})\Psi =0.
\end{equation}
This equation of motion has been studied for a long time. The solution of this equation with definite frequency can be decomposed as \cite{Bardeen1972,Berti:2005gp,Hod:2015cqa,Brill:1972xj,Teukolsky:1972my,Teukolsky:1973ha}
\begin{equation}\label{wavef}
{{\Psi }}(t, r, \theta, \phi ) =\sum\limits_{l,m}R_{lm}(r)S_{lm}(\theta)e^{im\phi }e^{-i\omega t},
\end{equation}
where\ $l$\ is\ the\ spherical\ harmonic\ index,\ $m$\ is\ the\ azimuthal\ harmonic\ index\ with\
$-l\le m\le l$\ and\ $ \omega$\ is\ the\ angular frequency\ of\ the scalar mode. $S_{lm}(\theta)$ is the scalar spheroidal harmonics
 satisfying\ the\ angular equation of motion
\begin{equation}\label{angeq}
\frac{1}{\sin \theta }\frac{d}{d\theta }(\sin \theta \frac{d{{S}_{lm}}}{d\theta })+[\l_{lm}+{{a}^{2}}({{\mu }^{2}}-{{\omega }^{2}}){{\sin }^{2}}\theta -\frac{{{m}^{2}}}{{{\sin }^{2}}\theta }]{{S}_{lm}}=0,
\end{equation}
where $\l_{lm}$ is an angular eigenvalue\footnote{Compared with the similar parameter $_{0}A_{lm}$ in \cite{Berti:2005gp}, $\l_{lm}=$$_{0}A_{lm}-a^2(\mu^2-\o^2).$}. When $a^2(\mu^2-\omega^2)>0$, the eigenfunction $S_{lm}$ is called prolate; when $a^2(\mu^2-\omega^2)<0$, the eigenfunction $S_{lm}$ is called oblate.

Although the angular equation \eqref{angeq} has been studied for a long time, an explicitly analytic expression of the general eigenvalue $\l_{lm}$ is still lacking \cite{Berti:2005gp}. In various limit cases, we  know asymptotic analytic expressions for the eigenvalues. When $|a^2(\mu^2-\omega^2)|\ll 1$,
the leading order of the  angular eigenvalue for both prolate and oblate cases is $l(l+1)$. This is obvious since the angular equation is approaching to Legendre equation in this case. When
$
|a^2(\mu^2-\omega^2)|\gg 1,
$
 the asymptotic behaviors of prolate and oblate eigenvalues are remarkably different. The asymptotic eigenvalues for the prolate and oblate cases are respectively,
\bea
\l_{lm}\sim -a^2(\mu^2-\omega^2);~~~ \l_{lm}\sim 2 q_{lm} a\sqrt{\mu^2-\omega^2},
\eea
where $q_{lm}$ is some numerical constant \cite{Berti:2005gp}. In our discussion, the prolate case is important. According to the discussion in reference \cite{Berti:2005gp}, one key result about the prolate angular eigenvalue is
\bea\label{bzhzh}
\l_{lm}>l(l+1)-a^2(\mu^2-\o^2),
\eea
which is an important bound for our following discussion. Moreover, it is remarkable to mention that some useful and lightening discussion of the eigenvalue bound had also been presented in \cite{Bardeen1972}.

\section{The radial equation of motion and effective potential}
The\ radial equation of \eqref{KG} obeyed\ by\ ${{R}_{lm}}(r)$\ is\ given\ by
\begin{equation}\label{radeq}
\Delta \frac{d}{dr}(\Delta \frac{dR_{lm}}{dr})+UR_{lm}=0,
\end{equation}
where
\begin{equation}
U = {[\omega ({{r}^2} + {a^2}) - {{m}}a]^2} + \Delta [2{{m}}a\omega  - \mu^2(r^2 + {a^2}) - {\l_{lm}}].
\end{equation}
In\ order\ to\ study\ the\ superradiant\ modes\ of\ black\ holes\ to\ the massive\ perturbation,\ the\ asymptotic\ solutions\ of\ the\ radial\ equation\ near\ the\ horizon\ and\ infinity\ are\ considered\ under\ appropriate\ boundary\ conditions. Defining\ the\ tortoise\ coordinate\ ${{r}_{*}}$\  by\ equation\ $ \frac{d{{r}_{*}}}{d{{r}}}=\frac{{{r}^{2}+a^2}}{\Delta }$ \  and\ a new\ radial\ function\ $ \tilde{R}_{lm} =\sqrt{r^2+a^2}R_{lm}$, the above\ radial\ wave\ equation\ can\ be\ rewritten\ as

\begin{equation}
\frac{d^{2}\tilde{R}_{lm} }{dr_{*}^{2}}+\tilde{U}\tilde{R}_{lm} =0.
\end{equation}
The asymptotic behaviors of $\tilde{U}$ at $r_*\rightarrow \pm\infty$ are as follows:
\begin{eqnarray}
\tilde{U}&\rightarrow& (\o-m\Omega_H)^2,~~~r_*\rightarrow -\infty(r\rightarrow r_+);\\
\tilde{U}&\rightarrow& \o^2-\mu^2,~~~r_*\rightarrow +\infty(r\rightarrow +\infty).
\end{eqnarray}
The physical boundary conditions that we are interested in  are ingoing wave at the horizon(${{r}_{*}}\to -\infty $ )and bound states (exponentially decaying modes) at spatial infinity(${{r}_{*}}\to +\infty $). Then the asymptotic solutions of the radial wave equation are the following
\begin{equation}
r\to +\infty ({{r}_{*}}\to +\infty ),~~~ \tilde{R}_{lm}\sim {{e}^{-\sqrt{{{\mu }^{2}}-{{\omega }^{2}}}{{r}_{*}}}},
\end{equation}
\begin{equation}
r\to {{r}_{+}}({{r}_{*}}\to -\infty ),~~~\tilde{R}_{lm}\sim {{e}^{-i(\omega -m{{\Omega }_{H}}){{r}_{*}}}}.
\end{equation}
It is easy to see that in order to get the decaying modes we need following condition
\begin{equation}
{{\omega ^2}<{\mu ^2}}.
\end{equation}

In order to analyse the superradiant stability of the Kerr black hole, we define a new radial wavefunction $\varphi=\D^{1/2} R$. The radial equation \eqref{radeq} is transformed into a Schrodinger-like wave equation with effective potential ${{V}_{1}}$,
\begin{equation}\label{V1}
\frac{{{d^2}\varphi }}{{d{r^2}}} + ({\omega ^2} - {V_1})\varphi  = 0,~~
{V_1} = {\omega ^2} - \frac{{U + {M^2} - {a^2}}}{{{\Delta ^2}}}.
\end{equation}
In order to see if there exist a trapping potential outside the horizon, we should analyze the shape of the effective potential ${{V}_{1}}$. From the following asymptotic behaviors of the potential ${{V}_{1}}$:
\begin{equation}\label{asyp1}
{{V}_{1}}(r\to \infty )\to {{\mu }^{2}} - \frac{{4M{\omega ^2} - 2M{\mu ^2}}}{r}+O(\frac{1}{{{r}^{2}}}),
\end{equation}
\begin{equation}\label{asyp2}
{{V}_{1}}(r\to {{r}_{+}})\to -\infty ,{{V}_{1}}(r\to {{r}_{-}})\to -\infty ,
\end{equation}
\begin{equation}\label{asyp3}
{{V}_{1}}^{'}(r\to \infty )\to \frac{{4M{\omega ^2} - 2M{\mu ^2}}}{{{r}^{2}}}+O(\frac{1}{{{r}^{3}}}).
\end{equation}
So when
\begin{equation}\label{omegaMu}
{{\omega ^2}<{\mu ^2\over 2}},
\end{equation}then
\begin{equation}
{{V}_{1}}^{'}(r\to \infty )<0.
\end{equation}
This means that there is no potential wells when $ r\to \infty$ and the black hole may be superradiantly stable, so \eqref{omegaMu} is one important basic inequality.
In the next section, we will find the regions of the parameter space where there is only one extreme outside the event horizon $r_+$ for effective potential ${{V}_{1}}$, no trapping well exists, which is separated from the horizon by a potential barrier, and the Kerr black holes are superradiantly stable.

\section{{The superradiant stability analysis}}
In this section, we will determine the regions of the parameter space where the system of Kerr black hole and massive scalar is superradiantly stable.
We define a new variable $z$,  $z  =  r - {r_ - }$. Then the\ explicit\ expression\ of\ the\ derivative\ of\ the\ effective\ potential $V_1$ \ is
\be
V_1^{'}(r)={{V}_{1}}^{'} (z)= \frac{{A{r^4} + B{r^3} + C{r^2} + Dr + E}}{{ - {\Delta ^3}}} = \frac{{{A_1}{z^4} + {B_1}{z^3} + {C_1}{z^2} + {D_1}z + {E_1}}}{{ - {\Delta ^3}}};
\ee

\begin{equation}
{A_1} = A;{B_1} = B + 4{r_ - }{A_1};{C_1} = C + (3{r_ - }){B_1} + (6r_ - ^2){A_1};
\end{equation}
\begin{equation}
{D_1} = D + (4r_ - ^3){A_1} + (3r_ - ^2){B_1} + (2{r_ - }){C_1};
\end{equation}
\begin{equation}
{E_1} = E + r_ - ^4{A_1} + r_ - ^3{B_1} + r_ - ^2{C_1} + {r_ - }{D_1};
\end{equation}

\bea
{A_1}& =& 2M( {\mu ^2}-2{\omega ^2}),\\
B_1&=&-16M r_- \omega ^2-\mu ^2 r_+^2+2 \l_{lm}+3 \mu ^2 r_-^2+2 a^2 \mu ^2,\\
C_1&=&-24Mr_-^2 \omega^2 +6\omega a m (\rm+\rp)-3(r_+-r_-) \left(a^2 \mu ^2+\rm^2\mu ^2 +\l_{lm} \right),\\
D_1&=&-16Mr_-^3\omega^2+4a M m(5r_--r_+)\omega+(r_+-r_-)^2(\mu^2r_-^2+\l_{lm}-1)+a^2(\mu^2(r_+-r_-)^2-4m^2),\\
E_1&=& 4M({a^4} - r_ - ^4){\omega ^2}+8M m a(r_-^2-a^2)\o+2 (\rp-\rm) \left(M^2+m^2a^2-a^2\right).
\eea

We denote the numerator of the derivative of the effective potential $V_1$ by $f(z)$, which is a quartic polynomial in $z$. Whether there is a trapping well outside the
horizon can be analyzed through the property of roots of equation $f(z)=0$. The four roots for equation $f(z)=0$ are $z_1,z_2,z_3$  and $z_4$. According to Vieta theorem,
we have the following relations,

\bea\label{vita1}
z_1z_2z_3z_4&=&\frac{E_1}{A_1},\\\label{vita2}
z_1z_2z_3+z_1z_2z_4+z_1z_3z_4+z_2z_3z_4&=&-\frac{D_1}{A_1},\\\label{vita3}
z_1z_2+z_1z_3+z_1z_4+z_2z_3+z_2z_4+z_3z_4&=&\frac{C_1}{A_1},\\\label{vita4}
z_1+z_2+z_3+z_4&=&-\frac{B_1}{A_1}.
\eea

Based on the asymptotic behaviors of the effective potential at the inner and outer horizons and infinity \eqref{asyp1}\eqref{asyp2}, one can
deduce that equation $V'_1(z)=0(\text{or}~ f(z)=0)$ has at least two positive real roots when $z> 0$. These two positive roots are denoted by $z_1, z_2$, namely
\be
z_1>0, \quad \quad z_2>0.
\ee
In the following, we will find a parameter region of the system where there are only  two positive roots for the equation $V'_1(z)=0$, i.e., $z_3,z_4$ are both negative. Then there is no trapping potential well acting as a mirror outside the horizon of the Kerr black hole. The system is superradiantly stable in this region.

Under the condition \eqref{omegaMu}, ${{{A}_{1}}}> 0$. Taking $E_1$ as a quadratic function of $\o$, one can show that $E_1>0$ without subtlety.  $E_1$ can be written as follows,
\begin{equation}
{E_1}(\o) = 4M({a^4} - r_ - ^4){\omega ^2} +8M m a(r_-^2-a^2)\o+2 (\rp-\rm) \left(M^2+m^2a^2-a^2\right).
\end{equation}
The discriminant of $E_1$ can be calculated directly as following,
\bea
\Delta_{E_1}=-4r_-^2(r_+-r_-)^4(r_++r_-)^2,
 \eea
which is obviously negative, $\Delta_{E_1}<0$ . The coefficient of $\omega^2$ in $E_1$ is obviously larger than $0$.  One can conclude that $E_1>0$ for any $\omega$.
Then according to \eqref{vita1}, we have
\be \label{4z1}
z_1 z_2 z_3 z_4>0.
\ee
As we mentioned above, $z_1$ and $z_2$ have been denoted as two positive roots of the derivative of the effective potential. From the above equation, we find that $z_3$ and $z_4$ can only be both positive or both negative, if they are real roots.

Next we turn to focus on the following equation in detail,
 \bea
 z_1+z_2+z_3+z_4&=&-\frac{B_1}{A_1}.
 \eea
 Because $A_1>0$, we want to identify a parameter region where $B_1>0$, then the roots $z_3, z_4$ are both negative.  Using the condition of the eigenvalue of our angular equation given in \eqref{bzhzh}, we have
\be
B_1>-(6r_+ r_- + 8 r_-^2)\omega^2+2l(l+1)+\mu^2(3r_-^2-r_+^2).
\ee
Now we define a quadratic function with respect to $\omega$ as follows:
\be
g(\omega)=-(6r_+ r_- + 8r_-^2)\omega^2+2l(l+1)+\mu^2(3r_-^2-r_+^2).
\ee
Identifying a parameter region where $B_1>0$ is transformed into identifying the parameter region where $g(\o)>0$. We will do  this in the following.

As mentioned in the introduction, we focus on discussing  Kerr black holes and massive scalar perturbation system in the regime $\mu<\sqrt{2}m \Omega_{H}$, i.e., $\mu^2<\frac{2 m^2 a^2}{(r^2+a^2)^2}$.
It is easy to see that if $g(\omega)>0$ is true for some real $\o$, the following inequality should be satisfied first,
\be
2l(l+1)+\mu^2(3r_-^2-r_+^2)>0.
\ee
 Using the conditions $l(l+1)>m^2$ and $\mu^2<\frac{2m^2 a^2}{(r^2+a^2)^2}$,  we  have
\be
2l(l+1)-\mu^2 r_+^2+3\mu^2 r_-^2>2m^2-\mu^2 r_+^2+3\mu^2 r_-^2>\mu^2 r_+^2 \Big(\frac{(r_+^2+ a^2)^2}{r_- r_+^3}-1\Big)+3\mu^2 r_-^2.
\ee
It is worth to pointing out that
\be
\frac{(r_+^2+ a^2)^2}{r_- r_+^3}=(1+\frac{r_-}{r_+})(1+\frac{r_+}{r_-})>1.
\ee
So we conclude that the following inequality
\be
2l(l+1)-\mu^2 r_+^2+3\mu^2 r_-^2>0,
\ee
could be always satisfied in our Kerr black hole and massive scalar perturbation system.

From our discussion above we have known that there must exist two real roots for $g(\omega)$ with respect to $\o$.
Taking $\omega_0$ as the positive real root of $g(\omega)$, then we have
\be
\omega_0^2=\frac{2l(l+1)+\mu^2(3r_-^2-r_+^2)}{8r_-^2+6 r_- r_+}.
\ee
When $0<\o<\o_0$, one can obtain $g(\o)>0$. Remember that in our system the parameters of massive scalar perturbation satisfy \eqref{omegaMu}. So we will have $g(\omega)>0$, if $\omega_0^2>\frac{\mu^2}{2}$, i.e.,
\be\label{ineq1}
\frac{2l(l+1)+\mu^2(3r_-^2-r_+^2)}{8r_-^2+6 r_- r_+}-\frac{\mu^2}{2}>0.
\ee
 Considering the relations $\mu <\sqrt{2}m\Omega_H$ and $l(l+1)>m^2$, we find
 \be
 2l(l+1)>2m^2>\frac{\mu^2}{\Omega_H^2}.
 \ee
Then, the previous inequality \eqref{ineq1} will hold if the following inequality holds
\be
\mu^2\Big(\frac{\frac{1}{\Omega_H^2}+(3r_-^2+r_+^2)}{8r_- + 6 r_- r_+}-\frac{1}{2}\Big)=\mu^2\Big(\frac{\frac{(r_+^2+r_- r_+)^2}{r_- r_+}+(3r_-^2+r_+^2)}{8r_- + 6 r_- r_+}-\frac{1}{2}\Big)>0,
\ee
where we use that $\Omega_H=\frac{\sqrt{r_+ r_-}}{r_+^2+r_- r_+}$. After some reorganization, we can obtain
\be
\frac{\mu^2(r_+^3+r_+ r_-^2-2 r_-^2 r_+ - r_-^3)}{2 r_-^2(4 r_- + 3 r_+)}>0.
\ee
Since $\frac{\mu^2 r_+^3 }{2 r_-^2(4 r_- +3 r_+)}>0$,  the above inequality is equivalent to the following one
\be \label{lam}
1+\lambda-2\lambda^2-\lambda^3>0,
\ee
where $\lambda=\frac{r_-}{r_+}\in (0,1]$, which is only related to the  parameters of the Kerr black hole.

\begin{figure}[htp]
\begin{center}
\includegraphics[width=200pt]{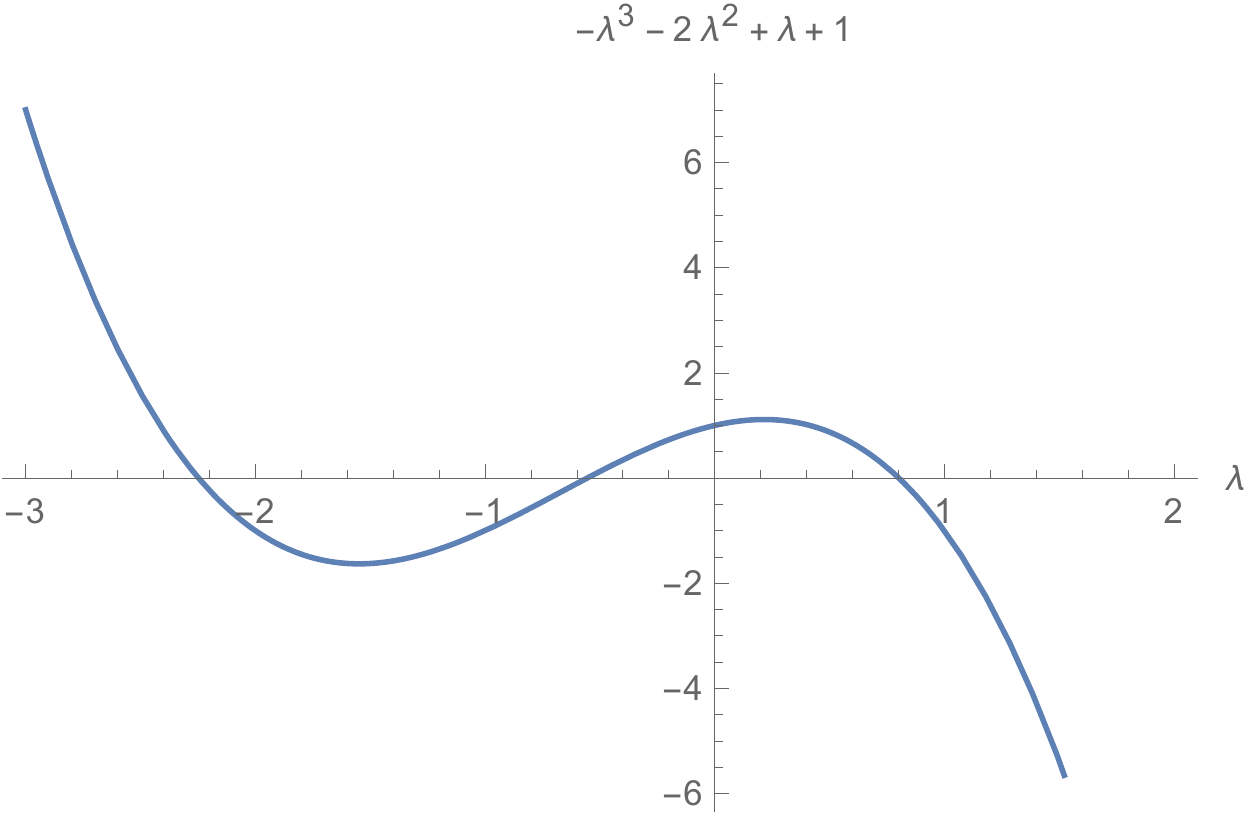}
\end{center}
\caption{Here we plot the curve of $1+\lambda-2\lambda^2-\lambda^3$ in the regime $\lambda\in [-3, 3/2]$. Actually we just need to focus on the interval  $0<\lambda<1$. From this curve one can obviously see that $1+\lambda-2\lambda^2-\lambda^3>0$ when $\lambda \in (0,\lambda_1)$.}
\label{figure1}
\end{figure}

There exist three real roots in the cubic equation $1+\lambda-2\lambda^2-\lambda^3=0$ and here we list them as
\bea
&&\lambda_1=\frac{1}{3}\Big(-2+2\sqrt{7}\cos \frac{\theta}{3} \Big),
\cr &&
\lambda_2=\frac{1}{3}\Big(-2-\sqrt{7}\Big(\cos \frac{\theta}{3}+\sqrt{3}\sin \frac{\theta}{3}\Big)\Big),
\cr &&
\lambda_3=\frac{1}{3}\Big(-2-\sqrt{7}\Big(\cos \frac{\theta}{3}-\sqrt{3}\sin \frac{\theta}{3}\Big)\Big),
\eea
where $\theta=\arccos (-\frac{1}{2\sqrt{7}})$. Note that $\lambda_1\approx 0.802>0,\, \lambda_2=-0.555<0, \, \lambda_3=-2.247<0$. One could plot the curve  for the cubic function $1+\lambda-2\lambda^2-\lambda^3$ with respect to $\lambda$, see Fig.(\ref{figure1}),  from which one can see  $1+\lambda-2\lambda^2-\lambda^3>0$ when $\lambda \in (0,\lambda_1)$.

So we prove that when the parameters of the Kerr black hole satisfy the following inequality,
\be
0<\frac{r_-}{r_+}<\frac{1}{3}\Big(-2+2\sqrt{7}\cos \frac{\theta}{3} \Big)\, , \quad \quad  \theta=\arccos (-\frac{1}{2\sqrt{7}}),
\ee
we have $B_1>g(\o)>0$. Then there is no trapping potential well for the effective potential of the radial equation of motion and the
Kerr black hole and scalar perturbation system is superradiantly stable.

\section{{Conclusion}}
In the present paper, we investigate the superradiant stability of a system with a Kerr black hole and massive scalar perturbation. Our
discussion is focusing on models where the mass scalar perturbation is smaller than an upper bound related with the angular velocity of the Kerr black hole horizon,
i.e., $\mu<\sqrt{2}m \Omega_H$, and is a complementary study of previous work.

 The equation of motion of the scalar perturbation in the Kerr black hole background is separated into angular and radial parts.
In our discussion, the spheroidal angular equation is prolate and we choose a general bound for the eigenvalue of this equation. The radial equation can be transformed
into a Schrodinger-like equation and the effective potential is important for the stability analysis. We find that when the parameters of scalar perturbation satisfy
$\omega<\frac{\mu}{\sqrt{2}}$ and the parameters of Kerr black hole satisfy $\frac{r_-}{r_+}<0.802$, there is no potential well outside the black hole horizon acting as
a mirror and the system is superradiantly stable.

Our study in this paper treats the Kerr black hole as a background geometry and only considers the dynamics of the free scalar perturbation.
In order to get a more refined result of the superradiant behavior of the Kerr black hole and scalar perturbation system, one further step is
 to study the coupled nonlinear equations of motion of the scalar perturbation and black hole.
 Recently, such  nonlinear evolution of superradiant process  has been studied in several models \cite{Sanchis-Gual:2015lje,Bosch:2016vcp,East:2017ovw,Chesler:2018txn,Cardoso:jcap}, and has found some exciting applications in astrophysical physics.
Another interesting extension is to study a model with nonlinear self-interacting scalar perturbation. It is  pointed
out that when we take self-interaction into account for the scalar perturbation, superradiant behavior of the system will be different and
non-linear scalar hairs will exist\cite{Hong:2019mcj}.
It will be interesting to  investigate the detailed superradiant behavior of a system consisting of a Kerr black hole and a scalar field with self-interaction.

{\bf Acknowledgements:}\\

Z.-F.Mai thanks Professor H. L¡§u for useful discussion. J.-H. Huang is supported by the Natural Science Foundation of
Guangdong Province (No.2016A030313444).

\end{document}